\documentclass[conference]{IEEEtran}

\IEEEoverridecommandlockouts
\usepackage{cite}
\usepackage{amsmath,amssymb,amsfonts}
\usepackage{float}

\usepackage{algorithm}
\IEEEoverridecommandlockouts

\usepackage{algorithmicx}
\usepackage{algpseudocode}
\usepackage{graphicx}
\usepackage{tikz}
\usepackage{textcomp}
\usepackage{xcolor}
\def\BibTeX{{\rm B\kern-.05em{\sc i\kern-.025em b}\kern-.08em
    T\kern-.1667em\lower.7ex\hbox{E}\kern-.125emX}}
\begin{document}

\title{ GCN-Driven Reinforcement Learning for Probabilistic Real-Time Guarantees in Industrial URLLC\\

{\footnotesize \textsuperscript{}}
\thanks{}
}

\author{
    \IEEEauthorblockN{Eman Alqudah, Ashfaq Khokhar}
    \IEEEauthorblockA{
        \textit{Department of Electrical and Computer Engineering}\\
        \textit{Iowa State University, USA}\\
        \{alqudah, ashfaq\}@iastate.edu
    }

}

\maketitle
\IEEEpubid{\makebox[\columnwidth]{%
  © 2025 IEEE. Personal use is permitted. Permission required for reuse.\hfill}%
  \hspace{\columnsep}\makebox[\columnwidth]{}
}

\begin{abstract}
Ensuring packet-level communication quality is vital for ultra-reliable, low-latency communications (URLLC) in large-scale industrial wireless networks. We enhance the Local Deadline Partition (LDP) algorithm by introducing a Graph Convolutional Network (GCN) integrated with a Deep Q-Network (DQN) reinforcement learning framework for improved interference coordination in multi-cell, multi-channel networks. Unlike LDP's static priorities, our approach dynamically learns link priorities based on real-time traffic demand, network topology, remaining transmission opportunities, and interference patterns. The GCN captures spatial dependencies, while the DQN enables adaptive scheduling decisions through reward-guided exploration. Simulation results show that our GCN-DQN model achieves mean SINR improvements of 179.6\%, 197.4\%, and 175.2\% over LDP across three network configurations. Additionally, the GCN-DQN model demonstrates mean SINR improvements of 31.5\%, 53.0\%, and 84.7\% over our previous CNN-based approach across the same configurations. These results underscore the effectiveness of our GCN-DQN model in addressing complex URLLC requirements with minimal overhead and superior network performance.
\end{abstract}


\section{Introduction}
Industrial ultra-reliable low-latency communications (URLLC), fueled by 5G+ and emerging technologies, are poised to significantly boost the performance, flexibility, and robustness of industrial cyber-physical systems (CPS). These systems must operate under stringent timing constraints in mission-critical tasks such as real-time sensing, robotic control, automated processes, and power grid operations \cite{b1}. In Extended Reality (XR) environments, reliable and low-latency communication is vital to ensure smooth 3D scene rendering, as any delay or packet loss can negatively impact the user experience \cite{b2}. Likewise, in industrial control systems, communication failures may jeopardize operational safety and stability. However, enabling real-time communication across multi-cell industrial wireless networks remains a major challenge, largely due to inherent transmission delays and processing overheads \cite{b3}.

\subsection{Related Work}
Real-time communication in wireless networks has been widely studied using both optimization-based approaches~\cite{b4,b5} and learning-based formulations~\cite{b6,b7}. These methods have been applied to resource allocation and scheduling problems, including classic strategies such as earliest-deadline-first (EDF) and rate-monotonic (RM) scheduling~\cite{b3,b8}. Additionally, other studies have focused on long-term performance metrics such as mean delay and age-of-information (AoI)~\cite{b9,b10}. Despite these advancements, achieving reliable and low-latency communication, especially in multi-cell environments, remains a significant challenge. Recent efforts, such as 5G configured grant (CG) scheduling~\cite{b11,b12}, have shown promise for real-time communication but are limited to uplink transmissions and lack flexibility for broader system applications.

A more recent approach, Local Deadline Partitioning (LDP)~\cite{b13}, has shown potential for improving real-time task scheduling in multi-cell wireless systems. By partitioning tasks according to local deadlines, LDP aims to meet stringent latency requirements under varying resource constraints. However, LDP's reliance on static priority assignments makes it less adaptable to dynamic network conditions, and its computational complexity grows with system size, which raises scalability concerns. To address these limitations, a CNN-based enhancement to LDP was proposed in our recent work~\cite{b14}, where a Convolutional Neural Network and graph coloring techniques dynamically predict link priorities based on real-time traffic and network states. This approach demonstrated significant improvements in resource allocation efficiency, SINR, and overall network schedulability, particularly in complex URLLC scenarios.

Graph-based deep learning methods, particularly Graph Convolutional Networks (GCNs) combined with Deep Q-Networks (DQNs), offer a promising solution to resource allocation in wireless networks\cite{b15}. GCNs are capable of capturing both the topological structure and dynamic interactions among network nodes, while DQNs allow for adaptive, reward-driven decision-making. In this paper, we extend the LDP framework by integrating a GCN-DQN-based decision-making model. This model learns optimal link selection policies based on link priority, interference, and traffic urgency. The GCN component encodes spatial dependencies using the conflict graph, while the DQN agent exploits this representation to make scheduling decisions that maximize long-term performance. By leveraging the graph structure and adapting to changing network conditions, our approach enables scalable, interference-aware scheduling with real-time guarantees, offering a robust solution for industrial URLLC systems operating in multi-cell environments.

To address the challenges of efficient resource block allocation, we propose a hybrid GCN-DQN framework that combines the spatial modeling capabilities of GCNs with the dynamic decision-making power of DQNs. This approach offers several advantages: it provides topology-aware scheduling by learning interference patterns across multi-cell networks, dynamically optimizes resource allocation in response to varying traffic loads, and scales efficiently to large network sizes. The real-time optimization capabilities of the GCN-DQN framework make it particularly suitable for industrial URLLC applications, where stringent timing and reliability constraints are crucial. By enhancing the LDP framework with GCN-based spatial modeling and DQN-driven policy learning, our work significantly improves adaptability, performance, and reliability in complex industrial wireless environments.

The rest of the paper is organized as follows. Section II presents the system model and formulates the resource allocation problem. Section III introduces the proposed solution, which integrates a GCN-assisted Deep Q-Network (DQN) reinforcement learning approach with a modified Local Deadline Partitioning (LDP) framework. Section IV describes the experimental setup and evaluation methodology, while Section V discusses the results and key performance insights. Finally, the conclusion offered in Section VI. 

\section{System Model and Problem Statement}
This paper addresses the scheduling problem in a wireless network by formulating it as an optimization task to efficiently allocate resources, considering link interactions and system constraints. We present a network model that captures key characteristics and interference dynamics.
\subsection{Network Model}

We consider a wireless network with \(N\) links, each representing a dedicated transmitter-receiver pair, operating over \(C\) orthogonal frequency channels. Each channel supports a maximum data capacity of \(B_c\) units per time slot. Time is slotted, and a subset of links is selected for transmission, subject to capacity and interference constraints.

The interference relationships among links are represented by a binary interference matrix \( \mathbf{I} \in \{0, 1\}^{N \times N} \), where \( \mathbf{I}_{ij} = 1 \) indicates that link \(j\) interferes with link \(i\), prohibiting simultaneous transmission on the same channel. This interference defines an undirected interference graph \( G = (V, E) \), where each vertex represents a link and an edge exists if the corresponding links interfere with each other.

Each link \(i\) is characterized by its traffic demand \( d_i \), deadline \( D_i \), period \( T_i \), signal-to-noise ratio (SNR), per-channel quality, and spatial coordinates \( (x_i, y_i) \), which model distance-based interference. We model physical interference using a log-distance path-loss model. The SINR at the receiver of link \( i \) is given by:

\[
\text{SINR}_i = \frac{P_t / d_i^{\alpha}}{\sum\limits_{j \in \mathcal{I}_i} P_t / d_{ji}^{\alpha} + N_0}  
\]

where \( P_t \) is the transmission power, \( \alpha \) is the path-loss exponent, \( d_i \) is the distance between the transmitter and receiver of link \(i\), and \( \mathcal{I}_i \) is the set of interfering links.

\subsection{Scheduling Problem}
Given the network model, the scheduling problem deals with allocating available frequencing channels, referred to as resource blocks (RBs), to links while maximizing SINR and satisfying real-time constraints, such as traffic demands and deadlines. Each link \( i \) has traffic demand \( d_i \), a deadline \( D_i \), and interference relationships with other links. The objective is to allocate RBs \( \{ r_i \} \) to links to maximize the overall SINR:

\[
\max_{r_i} \sum_{i \in V} \text{SINR}(r_i)
\]

where \( \text{SINR}(r_i) \) is the SINR for link \( i \), considering interference from other links and channel conditions. This problem is complex due to dynamic traffic, interference patterns, and the need to allocate resource blocks efficiently. To address these, we propose a solution based on a Graph Convolutional Network (GCN) model that predicts optimal RB allocations by capturing the underlying graph structure and interference relationships.

\section{Proposed GCN-RL Scheduling Framework}
This section introduces our GCN-based reinforcement learning (GCN-RL) framework for efficient wireless resource allocation. The objective is to optimize link scheduling for maximum SINR while accounting for traffic demand, interference, and deadline constraints.

Graph Convolutional Networks (GCNs) extract high-level features from a graph representing the wireless network, where nodes denote communication links and edges represent interference. A Deep Reinforcement Learning (DRL) agent then learns scheduling policies using these embeddings.

\subsection{Problem Formulation}
Consider a wireless network represented as a graph \( G = (V, E) \), where each node \( i \in V \) is a link, and an edge \( (i,j) \in E \) indicates interference. The goal is to assign resource blocks (RBs) to links under the following constraints:

\begin{itemize}
    \item \textbf{Resource Block Allocation:} Each link \( i \) is assigned one or more RBs \( rb_i \in \{0, 1, \dots, RB-1\} \), where \( RB \) is the total number of available RBs.
    
    \item \textbf{Interference Constraint:} Interfering links must not share the same RB:
    \[
    rb_i \neq rb_j \quad \forall (i, j) \in E
    \]
    
    \item \textbf{Traffic Demand Constraint:} Link \( i \)’s adjusted demand, accounting for interference from neighbors, must not exceed the RB capacity \( C \):
    \[
    \frac{d_i'}{1 + \sum_{j \in \mathcal{N}_i} M_{ij}} \leq C
    \]
    
    \item \textbf{Deadline Constraint:} Link \( i \)’s demand must be served within deadline \( D_i \):
    \[
    \frac{d_i'}{\text{effective demand}} \leq D_i
    \]
\end{itemize}

\subsection{Graph-Based Reformulation via MWIS}
The scheduling problem is reformulated as a Maximum Weighted Independent Set (MWIS) problem on the interference graph \( \mathcal{G} = (\mathcal{V}, \mathcal{E}) \), where each node has a weight \( w_i \) based on its demand, SINR, and deadline. The objective is:
\[
\max_{\mathcal{I} \subseteq \mathcal{V}} \sum_{v_i \in \mathcal{I}} w_i \quad \text{subject to} \quad (v_i, v_j) \notin \mathcal{E}, \forall v_i, v_j \in \mathcal{I}
\]
Due to MWIS being NP-hard, we use reinforcement learning to approximate an optimal solution through sequential decision-making.

\subsection{Formulation as a Markov Decision Process (MDP)}
The scheduling task is framed as an MDP with the following components:

\begin{itemize}
    \item \textbf{State (\(s_t\)):} Captures link-specific features ( SINR, demand, deadline) and the network’s structural context via GCN embeddings.
    
    \item \textbf{Action (\(a_t\)):} Chooses a link to schedule at time \( t \), ensuring the resulting set is independent.
    
    \item \textbf{Reward (\(r_t\)):} Reflects the gain from scheduling based on SINR:
    \[
    r_t = \log_2(1 + \text{SINR}_t)
    \]
    
    \item \textbf{Transition:} Represents changes in the system state after scheduling.
    
    \item \textbf{Policy (\(\pi(a_t | s_t)\)):} A neural policy network that maps states to actions to maximize expected cumulative rewards.
\end{itemize}

\subsection{GCN-Based Feature Aggregation}

A two-layer GCN is used to aggregate contextual and structural information. Each node \( v_i \) updates its feature representation as:
\[
\mathbf{h}_i^{(l+1)} = \sigma\left( \sum_{j \in \mathcal{N}(i) \cup \{i\}} \frac{1}{\sqrt{d_i d_j}} \mathbf{W}^{(l)} \mathbf{h}_j^{(l)} \right)
\]
where:
\begin{itemize}
    \item \(\mathbf{h}_i^{(l)}\): Node \(i\)'s feature vector at layer \(l\),
    \item \(d_i\): Degree of node \(i\),
    \item \(\mathbf{W}^{(l)}\): Trainable weight matrix,
    \item \(\sigma\): Nonlinear activation function (ReLU).
\end{itemize}

The final embeddings \(\mathbf{h}_i^{(L)}\) are mapped to Q-values:
\[
Q_i = \mathbf{w}^\top \mathbf{h}_i^{(L)} + b
\]
These guide the DRL agent in selecting links that are high-priority and low-interference.

\subsection{Training Procedure and Stability}

We employ strategies to ensure stable learning:

\begin{itemize}
    \item \textbf{Experience Replay:} Transitions \((s_t, a_t, r_t, s_{t+1})\) are stored and randomly sampled to break correlation.
    
    \item \textbf{Mini-Batch Updates:} A mean squared error loss is computed between predicted and target Q-values:
    \begin{align}
    \mathcal{L} &= \mathbb{E}_{(s, a, r, s') \sim \mathcal{D}} \left[ \left( Q(s, a) - y \right)^2 \right] \nonumber \\
    y &= r + \gamma \max_{a'} Q(s', a')
    \end{align}
\end{itemize}

\begin{figure}[H]
\centering
\includegraphics[width=\columnwidth,height=3.55in]{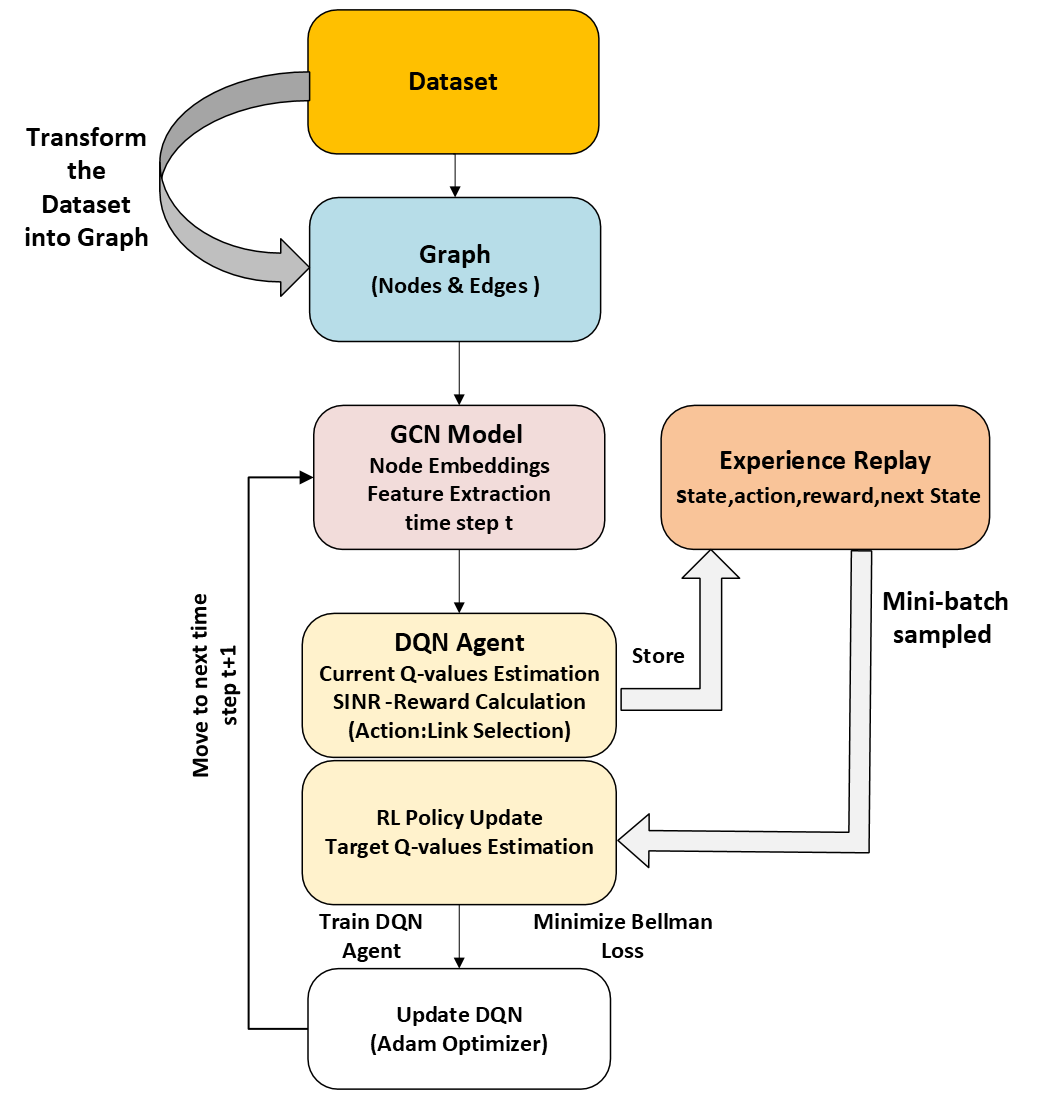}
\caption{Interaction between GCN and DQN Agent.}
\label{Interaction}
\end{figure}

Figure~\ref{Interaction} illustrates the interaction between the GCN and DQN components, forming the core of our framework: the GCN encodes the interference graph, the DQN estimates Q-values, and selected links update the policy.

\subsection{GCN-Based Modified Local-Deadline-Partition (LDP) Algorithm}
This work introduces a GCN-based enhancement of the Local-Deadline-Partition (LDP) algorithm to improve resource allocation performance under interference and real-time scheduling constraints. By leveraging the GCN framework, the model learns link priorities and interference patterns directly from network data, providing a data-driven alternative to heuristic methods. This approach enables adaptive scheduling based on dynamic demands, interference, and deadlines.

Integration of GCN predictions with the LDP framework enables dynamic policy updates in response to network variations. The detailed steps of the approach are shown in Algorithm~1.
bining graph-convolutional networ
{\small
\begin{algorithm}
\caption{Modified LDP Algorithm with GCN-DQN RL}
\begin{algorithmic}[1]
\State \textbf{Input:} $A_{i,1}$: Arrival time of first packet on link $i$; $\mathcal{M}_i$: Interfering links of $i$; $T_l, D_l$: Period and deadline of $l \in \mathcal{M}_i \cup \{i\}$; $X_{i,t}$: Local traffic demand; $\text{State}.l.rb.t$: Transmission state for $l$ and all $rb$ at $t$; Graph representation (nodes = links, edges = conflicts); Pretrained GCN and RL agent
\State \textbf{Output:} Optimized scheduling decisions using GCN + RL
\vspace{1mm}
\State \textbf{Step 1: Graph Construction}
\State Construct conflict graph with nodes (links) and edges (interference); Encode graph features using \textbf{GCN}
\vspace{1mm}
\State \textbf{Step 2: GCN-based Priority Computation}
\State Compute priority: $P_{\text{prio},i,t} = f_{\text{GCN}}(X_{i,t}, d''_{i,t}, \text{State}, \text{Graph})$
\State Share $P_{\text{prio},i,t}$ with interfering links in $\mathcal{M}_i$
\vspace{1mm}
\State \textbf{Step 3: RL Scheduling Decision using DQN}
\State Initialize \texttt{done} = \textbf{false}
\While{\texttt{done} == \textbf{false}}
  \State \texttt{done} = \textbf{true}
  \For{each $rb \in RB$ in parallel}
    \Statex \textbf{State Representation:} $s_i = (P_{\text{prio},i,t}, X_{i,t}, \text{State}, \text{Graph})$
    \State Predict Q-values: $Q(s_i, a)$ using DQN (actions = ACTIVE / INACTIVE)
    \If{Exploration (epsilon-greedy)}
      \State Choose $a = \arg\max_a Q(s_i, a)$ with exploration probability
    \Else
      \State Choose $a = \arg\max_a Q(s_i, a)$
    \EndIf
    \If{$a$ == ACTIVE}
      \State Assign $rb$ to link $i$; Update $X_{i,t} = X_{i,t} - 1$
    \Else
      \State Mark $rb$ as INACTIVE for link $i$
    \EndIf
    \State \textbf{Reward Calculation:} $r = \text{compute\_reward}(\text{State}, P_{\text{prio}}, D)$
    \State \textbf{DQN Update:} Store $(s_i, a, r, s_{i+1})$ in replay buffer
    \State Sample batch and update Q-network by minimizing:
    \[
      L(\theta) = \mathbb{E} \left[ \left( r + \gamma \max_{a'} Q(s_{i+1}, a'; \theta^-) - Q(s_i, a; \theta) \right)^2 \right]
    \]
  \EndFor
  \If{All RBs are decided}
    \State \texttt{done} = \textbf{true}
  \EndIf
\EndWhile
\vspace{1mm}
\State \textbf{Step 4: RL Environment Update}
\State Update environment; Compute new reward: $r = \text{compute\_reward}(\text{State}, P_{\text{prio}}, D)$
\State Store $(\text{state}, \text{action}, r, \text{next\_state})$ in RL memory
\vspace{1mm}
\State \textbf{Step 5: RL Policy Update}
\State Train RL agent (off-policy); Update \textbf{GCN} parameters to improve priority prediction
\vspace{1mm}
\State \textbf{Step 6: Continue Iteration Until Convergence}
\State Repeat for next time step $t+1$
\end{algorithmic}
\end{algorithm}
}

The proposed GCN-DQN based LDP algorithm enables real-time link scheduling under interference and deadline constraints by comks (GCNs) for link priority estimation and deep Q-Network (DQN) reinforcement learning for intelligent resource allocation.

\textbf{Step 1:} Conflict Graph Construction. At each time slot, the system constructs a conflict graph \( G = (\mathcal{V}, \mathcal{E}) \), where each node \( v_i \in \mathcal{V} \) represents a wireless link and each edge \( (v_i, v_j) \in \mathcal{E} \) denotes interference between links \( i \) and \( j \), derived from a predefined interference matrix. Each node is associated with feature vectors including traffic demand \( X_{i,t} \), remaining time to deadline \( d''_{i,t} \), and per-channel quality metrics.

\textbf{Step 2:} Link Priority Estimation via GCN. The GCN takes the graph and node features as input and computes priority scores for each link by capturing both local traffic characteristics and the broader interference topology. The link priority at time \( t \) is computed as:
\[
\text{Prio}_{i,t} = f_{\text{GCN}}(X_{i,t}, d''_{i,t}, \text{State}_{i,t}, G)
\]
Here, \( \text{State}_{i,t} \) encodes the local transmission state of link \( i \) and the current states of its neighbors. This priority acts as part of the DQN state input.

\textbf{Step 3:} Scheduling via DQN Agent. For each resource block (RB), a DQN agent selects an action for each link. The state observed by the agent includes the GCN-computed priority, current traffic status, deadline urgency, and RB availability. Using an $\varepsilon$-greedy policy, the agent chooses either to transmit (mark the RB as ``ACTIVE'') or defer (mark it as ``INACTIVE'') by evaluating Q-values. The agent avoids RB collisions by considering neighboring links' decisions through shared states.

\textbf{Step 4:} Environment and Traffic Update. After each RB scheduling decision, the environment updates the remaining traffic for each link and resolves any conflicts. Interference is checked to ensure only non-colliding transmissions proceed. Links with completed demands or expired deadlines are marked accordingly, affecting future states.

\textbf{Step 5:} Reward Calculation and Agent Update. A reward is computed based on several factors:
\begin{itemize}
    \item \textbf{Signal Quality:} Whether the SINR exceeds the required threshold for reliable transmission.
    \item \textbf{Timeliness:} Whether the link meets its deadline.
    \item \textbf{Efficiency:} Amount of traffic transmitted without conflict.
\end{itemize}
The experience tuple \( (s, a, r, s') \) is stored in a replay buffer. The agent periodically samples batches to update its policy using the Bellman equation:
\[
L(\theta) = \mathbb{E}\left[\left( r + \gamma \max_{a'} Q(s', a'; \theta^-) - Q(s, a; \theta) \right)^2\right]
\]
where \( \theta \) are the current network parameters and \( \theta^- \) are those of the target network, a copy of the Q-network that is updated less frequently to stabilize training.

The GCN is periodically refined with updated state representations as the environment evolves.

\textbf{Step 6:} Iterative Online Execution. This process repeats over time slots, enabling the system to adaptively schedule transmissions under time-varying traffic and interference conditions.

Compared to classical LDP approaches which use static priority rules or heuristic methods, the proposed GCN-DQN LDP algorithm is fully \textbf{adaptive}, \textbf{learning-based}, and \textbf{interference-aware}. It jointly optimizes link scheduling decisions by leveraging graph-based priority modeling and reinforcement learning-based action selection, significantly enhancing resource utilization and deadline satisfaction in dense wireless networks.

\subsection{GCN-DQN Architecture for Priority Prediction and Resource Block Selection}
Unlike traditional CNN models, the Graph Convolutional Network (GCN) processes graph-structured input, ideal for capturing interference relationships among communication links. The framework integrates link features and topology into a reinforcement learning pipeline, trained to maximize signal quality under interference and capacity constraints.

\subsubsection{Dataset}
The dataset contains 1000 time-step snapshots of a network with \( N \) links and \( C \) orthogonal channels. Features for each link include work density, signal-to-noise ratio (SNR), traffic demand, deadline, period, spatial position, and per-channel quality, concatenated into a L-dimensional vector where $L$ is the number of features added to the number of channels  per link. The dataset also includes a precomputed interference matrix and link position data for graph construction.
\subsubsection{Priority Prediction with GCN-DQN}  
The GCN-DQN model predicts link priorities using Q-values. It consists of two graph convolutional layers, each followed by ReLU activation:
\[
\text{GCNConv}_{1}: \mathbb{R}^{L} \rightarrow \mathbb{R}^{128}, \quad
\text{GCNConv}_{2}: \mathbb{R}^{128} \rightarrow \mathbb{R}^{128}
\]
The final Q-value for each link is produced by a linear output layer:
\[
\text{Q}_{i} = \text{Linear}(128) \in \mathbb{R}
\]
These Q-values are used to rank links, guiding the resource block (RB) allocation decisions.

\subsubsection{Resource Block Allocation Strategy}  
A greedy algorithm is employed to assign RBs based on the predicted Q-values and the link's traffic demand. The algorithm selects the highest Q-valued links and assigns them to the first available channel that satisfies the link's demand, within the channel capacity constraints.

\subsubsection{Optimization and Loss Functions}  
The GCN-DQN model is trained using Deep Q-Learning with a replay memory of size 10,000, which stores past transitions to enable randomized experience replay. This mechanism breaks correlations between sequential experiences and enhances training stability. At each training step, a mini-batch of size 32 is sampled from the replay buffer.

The loss function is the Mean Squared Error (MSE) between predicted and target Q-values. The target Q-values are computed using the Bellman equation, incorporating the reward and the maximum estimated Q-value of the next state. Training employs the Adam optimizer with a learning rate of 0.001 and a discount factor \( \gamma = 0.99 \).

\subsubsection{Pre-trained Model Outputs}  
Once trained, the GCN-DQN model provides real-time priority scores (Q-values) for each link, facilitating dynamic RB allocation. This approach enhances network performance by improving link-level signal quality and resource utilization, demonstrating robust performance under various interference conditions and supporting scalable scheduling in dynamic environments.
\begin{figure*}[!t]
    \centering
    \includegraphics[width=\textwidth, height=2.2in]{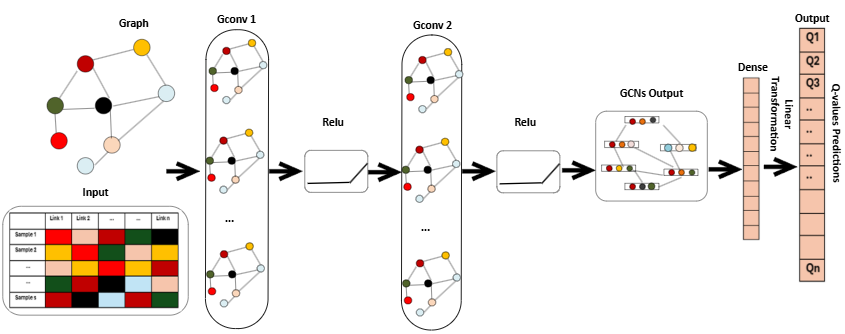} 
    \caption{Architecture of GCN-DQN Model.}
    \label{fig:model}
\end{figure*}

\section{Experimental Study}
In this work, we evaluate the performance of the proposed GCN-DQN-based link scheduling algorithm, which integrates graph convolutional networks and deep reinforcement learning to optimize resource allocation in multi-link industrial wireless networks. The algorithm leverages spatial and topological features to predict link priorities based on real-time traffic demand, remaining deadlines, and interference relationships. 

Resource block (RB) allocations are performed in a distributed and interference-aware manner, aiming to maximize signal quality (SINR) while satisfying traffic and latency constraints. The experimental study assesses the effectiveness and scalability of the proposed approach under varying traffic loads, interference patterns, and channel conditions. Results demonstrate that the model provides robust scheduling decisions with improved spectral efficiency and interference mitigation.

\subsection{Network and PPRC Traffic Settings}
We consider three networks of different sizes to represent various real-time network scenarios. The network size, number of channels, link/node spatial distribution density, and number of conflicting links per link are chosen based on industrial URLLC settings.

\begin{itemize}
    \item \textbf{Network 1}: 91 wireless nodes are deployed in a $120 \times 120$ square-meter region, generating 83 links. The network is organized into nine cells (3×3 grid), each with a base station (BS).
    \item \textbf{Network 2}: 151 wireless nodes are deployed in a $120 \times 120$ square-meter region, generating 163 links. The same 3×3 cell grid structure applies.
    \item \textbf{Network 3}: 320 wireless nodes are deployed in a $240 \times 240$ square-meter region, generating 324 links. The network consists of 36 cells (6×6 grid).
\end{itemize}

We apply the Wireless Industrial Indoor path loss model to determine the interference effect among links. Regarding channel allocation, we assume a 5G Numerology 4 setting where each resource block (RB) occupies 2.8 MHz. With a total 20 MHz bandwidth, there are 7 RBs available ($N=7$). To simulate diverse industrial URLLC scenarios, the available channels vary from 3 to 11.
\begin{table}[h]
    \centering
    \caption{Network Settings}
    \begin{tabular}{|l|c|}
        \hline
        \textbf{Parameter} & \textbf{Value} \\
        \hline
        Network Size & 120 × 120 m\textsuperscript{2}, 240 × 240 m\textsuperscript{2} \\
        \hline
        Number of Links & 83, 151, 320 \\
        \hline
        Channel Model & Indoor path loss model (Path loss coefficient = 3) \\
        \hline
        Bandwidth & 20 MHz \\
        \hline
        Number of Channels & 3–11 \\
        \hline
        Modulation & 16QAM \\
        \hline
        SINR Threshold & 15 dB \\
        \hline
        
    \end{tabular}
    \label{tab:network_settings}
\end{table}

\subsection{System Evaluation and Performance Metrics}
To assess the effectiveness of the proposed CNN-based link scheduling algorithm, we evaluate key performance metrics that reflect network efficiency, reliability, and schedulability.In addition ,we assesed the links qualities using SINR .

\subsubsection{Schedulability Condition and Schedulable Ratio}
We incorporate traffic demand as a key criterion for determining a link’s schedulability.A link is considered schedulable if its traffic demand can be accommodated within its allocated capacity before the deadline. Mathematically, for a given link \( i \), if the ratio of its traffic demand \( X_i \) to its link capacity \( C_i \) is less than or equal to its relative deadline \( D_i \), then the link can be scheduled within its deadline:

\begin{equation}
\frac{X_i}{C_i} \leq D_i
\end{equation}

On the other hand, if the traffic demand exceeds the available capacity within the given deadline, we determine that the link cannot be scheduled. In this case, the required number of transmissions surpasses the link's available capacity before the deadline, leading to a scheduling failure.

The \textit{schedulable ratio} quantifies the number of successfully scheduled links without violating deadlines. It is defined as:

\begin{equation}
\text{Schedulable Ratio} = \frac{\sum_{i=1}^{N} \mathbb{1} \left( \frac{X_i}{C_i} \leq D_i \right)}{N}
\end{equation}

where \( \mathbb{1}(\cdot) \) is an indicator function that returns 1 if the condition \( \frac{X_i}{C_i} \leq D_i \) holds and 0 otherwise. The numerator counts the number of schedulable links, and the denominator represents the total number of links in the network.

A higher schedulable ratio indicates better resource allocation efficiency, ensuring that most links meet their real-time constraints without excessive retransmissions or packet loss.

\subsubsection{Signal-to-Interference-plus-Noise Ratio (SINR)}
To assess the quality of received signals in the presence of interference and noise, we measure the SINR, a key metric for evaluating link quality ,as follows :
\begin{equation}
    \text{SINR} = \frac{P_{signal}}{P_{interference} + P_{noise}}
\end{equation}

where $P_{signal}$ is the received signal power, $P_{interference}$ is the sum of interference power from other links sharing the same resource block, and $P_{noise}$ represents background noise power. SINR values determine link reliability and retransmission needs.

The received signal power for the \( i \)th link, denoted as \( P_{\text{signal},i} \), is determined by the transmitted power \( P_{\text{tx}} \), the distance \( d_i \) between the transmitter and receiver of the \( i \)th link, and the path loss exponent \( \alpha \). It is expressed as:

\begin{equation}
    P_{\text{signal},i} = \frac{P_{\text{tx}}}{d_i^\alpha}
\end{equation}

The transmitted power \( P_{\text{tx}} \) is consiederd constant for all nodes and path loss exponent \( \alpha \) characterizes the rate at which the signal attenuates with distance, typically ranging from 2 in free-space environments to values above 3 in obstructed or indoor scenarios. As the distance \( d_i \) increases, the received signal power decreases exponentially due to propagation losses.

The interference power experienced by the \( i \)th link, denoted as \( P_{\text{interference},i} \), is caused by other links assigned to the same resource block and that interfere with link \( i \) based on the interference graph \( G \). The set of interfering links for \( i \) is denoted as \( \mathcal{I}_i \), which consists of all links that share an edge with \( i \) in the interference graph. This formulation captures the impact of co-channel interference while respecting the interference graph structure.The interference power is thus given by:

\begin{equation}
    P_{\text{interference},i} = \sum_{\substack{j \in \mathcal{I}_i \\ \text{RB}_i = \text{RB}_j}} \frac{P_{\text{tx}}}{d_{i,j}^\alpha}
\end{equation}

where \( d_{i,j} \) represents the distance between the transmitter of link \( i \) and the transmitter of link \( j \), and \( \alpha \) is the path loss exponent. The term \( \mathcal{I}_i \) ensures that interference is only summed over links that are directly connected to link \( i \) in the interference graph, which accurately models the conflict constraints in the network. This formulation explicitly accounts for interference relationships defined by the conflict graph,
ensuring realistic modeling of co-channel interference, which degrades the Signal-to-Interference-plus-Noise Ratio (SINR) and affects link reliability.

The noise power \( P_{\text{noise}} \) in a communication system is the power associated with the thermal noise in the system. It can be calculated using the following formula:
\begin{equation}
    P_{\text{noise}} = k T B
\end{equation}

where:
- \( k \) is Boltzmann's constant, \( k = 1.38 \times 10^{-23} \, \text{J/K} \),
- \( T \) is the system temperature in Kelvin. A common value for communication networks is \( T = 290 \, \text{K} \), which is considered room temperature,
- \( B \) is the bandwidth of the system in Hertz (Hz). We used the same assumption as in the paper\cite{b13} and assumed a bandwidth \( B = 20 \, \text{MHz} \).

This formula calculates the thermal noise power in the system, which is dependent on the temperature and bandwidth of the system. The noise power is critical in determining the signal-to-noise ratio (SNR) and, consequently, the performance of the communication system.

\subsubsection{Link Reliability}
Link reliability is defined as the fraction of links achieving SINR above a predefined threshold ($\gamma_{th}$). Given a set of SINR values, the reliability score is computed as:
\begin{equation}
    R_{reliability} = \frac{\sum (\text{SINR} \geq \gamma_{th})}{N_{links}}
\end{equation}
where $\gamma_{th} = 15$ dB is chosen to match the assumption in the base line paper \cite{b13}

\subsubsection{Network Capacity Estimation}
The estimated network capacity is derived by considering the number of schedulable and reliable links:
\begin{equation}
    C_{network} = N_{links} \times R_{schedulable} \times R_{reliability}
\end{equation}
This metric reflects the effective number of links capable of successful transmission in a given time slot.

\section{Results and Discussion}
This section presents the evaluation and analysis of the performance of our proposed GCN based solution compared to the traditional LDP algorithm. The results are based on simulations conducted under various network configurations, focusing on key performance metrics such as receiver-side SINR, prediction accuracy, and resource block selection efficiency. First, we outline the simulation setup used to generate the results, followed by a detailed presentation of the performance of the GCN pre-trained model across different experiments. The subsequent subsections discuss the observed trends, the significance of the improvements, and the implications of our findings.

\subsection{Simulation Setup}
The simulation experiments were conducted using Google Colab, a cloud-based platform that provides computational resources for machine learning and data processing tasks. The code was executed in Python, utilizing libraries such as TensorFlow, PyTorch, NumPy, and SciPy for efficient implementation. Hardware acceleration, including GPU support, was enabled when necessary to optimize performance.

For local development and preprocessing, an Intel\textregistered~Xeon\textregistered~CPU E3-1240 v6 @ 3.70 GHz system with 16 GB of RAM was used. The machine operates on a 64-bit architecture without touch or pen input capabilities. However, all computationally intensive tasks were offloaded to Colab to leverage cloud-based resources, ensuring efficient execution of deep learning models and optimization algorithms.

\subsection{Performance of GCN-Based Model}
The training results of our proposed GCN-based model demonstrate strong performance in predicting optimal channel allocations. The model achieved a Train Mean Absolute Error (MAE) of 0.0996 and a Test MAE of 0.1183, reflecting a high level of accuracy in both training and generalization. In addition, the model consistently reached near-perfect accuracy levels, with 100.00\% training accuracy and 99.88\% test accuracy.

The loss curves observed during training show a consistent downward trend, confirming effective learning. In the early stages, the model experienced higher error rates, but these rapidly declined as training progressed. In later epochs, both training and validation losses had converged, indicating strong model convergence and minimal overfitting.

\begin{figure}[htbp]
    \centering
    \includegraphics[width=\columnwidth]{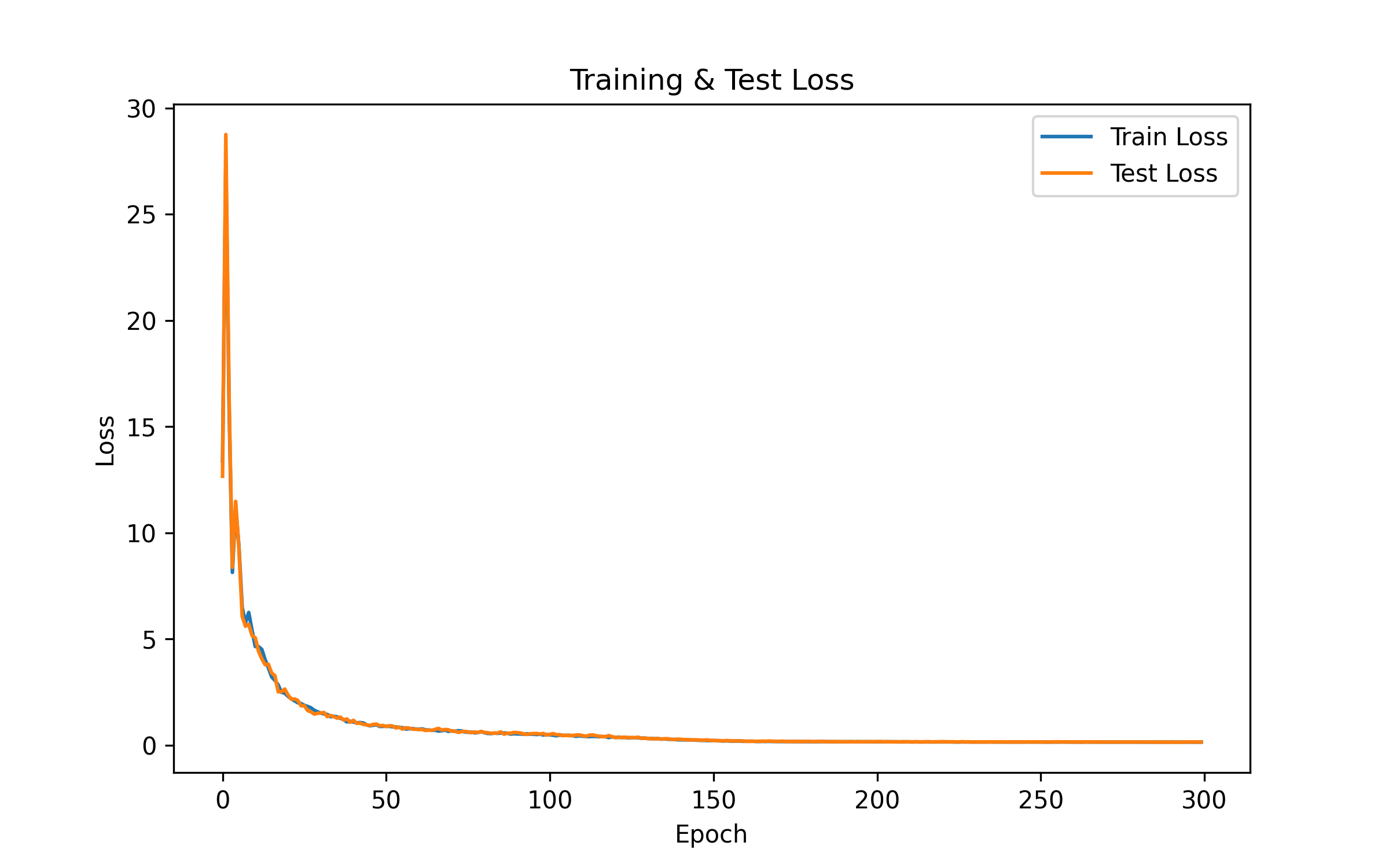} 
    \caption{Training and validation losses of the proposed GCN-based channel prediction model.}
    \label{Lossmodel}
\end{figure}

Figure~\ref{Lossmodel} presents the training and validation loss curves of the GCN model. The early sharp decline in loss illustrates the model's ability to quickly learn meaningful graph-structured features from the input data. Over subsequent epochs, the loss curves gradually stabilized at low values, demonstrating that the model effectively captured the spatial and relational dependencies among links in the network.

The close alignment between training and validation losses further confirms that the GCN model generalizes well to unseen data. Additionally, the minimal fluctuations in the loss curves highlight a well-regularized learning process, ensuring stability and robustness in performance across different dataset samples.

The steady reduction in loss and the stability of accuracy suggest that the chosen learning rate of $1.0 \times 10^{-2}$ was effective in optimizing the model. The minimal gap between training and validation loss further supports the claim that the model is well-fitted to the data and exhibits strong generalization capabilities.

\subsection{Receiver-Side SINR Performance}
To evaluate the impact of interference coordination on the performance of our proposed solution, we examined the receiver-side SINR across three distinct network configurations using three scheduling methods: the baseline LDP algorithm\cite{b13}, our previous CNN-based LDP algorithm \cite{b14}, and our proposed GCN-DQN approach.

In Network 1 (83 links), the baseline LDP algorithm achieved a mean SINR of 15.09 dB, with a 25\%-75\% range between 14.13 dB and 15.95 dB. The CNN-based method significantly outperformed LDP, achieving a mean SINR of 32.09 dB and a range from 31.28 dB to 32.88 dB. The highest SINR performance was achieved by our GCN-DQN approach, with a mean SINR of 42.20 dB and a range from 41.85 dB to 44.11 dB. These results demonstrate the effectiveness of deep reinforcement learning in managing interference.

In Network 2 (151 links), the LDP method recorded a mean SINR of 14.79 dB with a range from 13.94 dB to 16.12 dB, while the CNN-based approach achieved 28.75 dB with a range of 28.12 dB to 29.76 dB. Again, GCN-DQN delivered superior results, with a mean SINR of 43.99 dB and a narrower, higher-quality range between 43.77 dB and 44.33 dB, reflecting more consistent and reliable performance even in larger-scale settings.

In Network 3 (320 links), the LDP algorithm resulted in a mean SINR of 15.31 dB and a range from 14.03 dB to 16.59 dB. The CNN-based method improved upon this with a mean SINR of 22.81 dB and a range from 22.31 dB to 22.70 dB. The GCN-DQN approach once again outperformed the other methods, achieving a mean SINR of 42.14 dB with a narrow range of 41.98 dB to 42.10 dB. This highlights the robustness and scalability of our approach in highly dense environments, where interference coordination becomes more complex and critical. The significant SINR gain across all networks confirms that our GCN-DQN framework is well-suited for real-time link scheduling in 5G/6G networks, particularly in ultra-dense deployments where traditional methods struggle to deliver consistent performance.

These comparisons consistently highlight the effectiveness of the GCN-DQN scheduling strategy in boosting SINR, surpassing both the traditional LDP algorithm and the CNN-based method across various network sizes.

We also assessed the impact of receiver-side SINR on real-time schedulability. Our GCN-DQN algorithm demonstrated a consistent 100\% schedulability ratio across all network configurations, a significant improvement over the LDP algorithm. Specifically, while LDP achieved a 100\% schedulability ratio, our GCN-DQN based method attained this performance consistently across Network 1, Network 2, and Network 3. This indicates that our GCN-based interference mitigation strategy effectively minimizes interference to such an extent that all links become schedulable. This consistent 100\% schedulability underscores the superior real-time capacity and reliability offered by our GCN-based method, showcasing a substantial advancement over traditional interference management techniques. This demonstrates that our GCN-based approach, by achieving 100\% schedulability, significantly enhances the real-time capacity of the network, ensuring reliable URLLC communications.

\subsection{Comparison with CNN Model and Baseline LDP Approach}
To evaluate the performance of the proposed GCN-DQN model, we compare it with our previously implemented CNN-based scheduler and the baseline LDP method from paper~\cite{b13}. The comparison focuses on key performance metrics: mean SINR, interquartile range (25\%-75\%), SINR gain over the LDP baseline, percentage SINR improvement, and inference time. The results, for Network 1, are summarized in Table~\ref{tab:comparison}.

\begin{table}[H]
\centering
\caption{Comparison of Scheduling Methods}
\label{tab:comparison}
\scriptsize
\begin{tabular}{@{}|l|c|c|c|c|c|@{}}
\hline
\textbf{Method} & \textbf{Mean} & \textbf{Range} & \textbf{Gain} & \textbf{Improve.} & \textbf{Time} \\
                & SINR (dB)     & (dB)           & (dB)          & (\%)              & (s)           \\
\hline
LDP\cite{b13}        & 15.09  & [14.13, 15.95] & --     & --      & N/A     \\
CNN-based \cite{b14}  & 32.09  & [31.28, 32.88] & +16.99 & +112.5  & 0.47    \\
GCN-DQN    & 42.20  & [41.85, 44.11] & +27.11 & +179.6  & 0.0165  \\
\hline
\end{tabular}
\end{table}

\begin{figure}[!t]
\vspace{-5pt}
\centering
\includegraphics[width=\columnwidth ,height=2.0in]{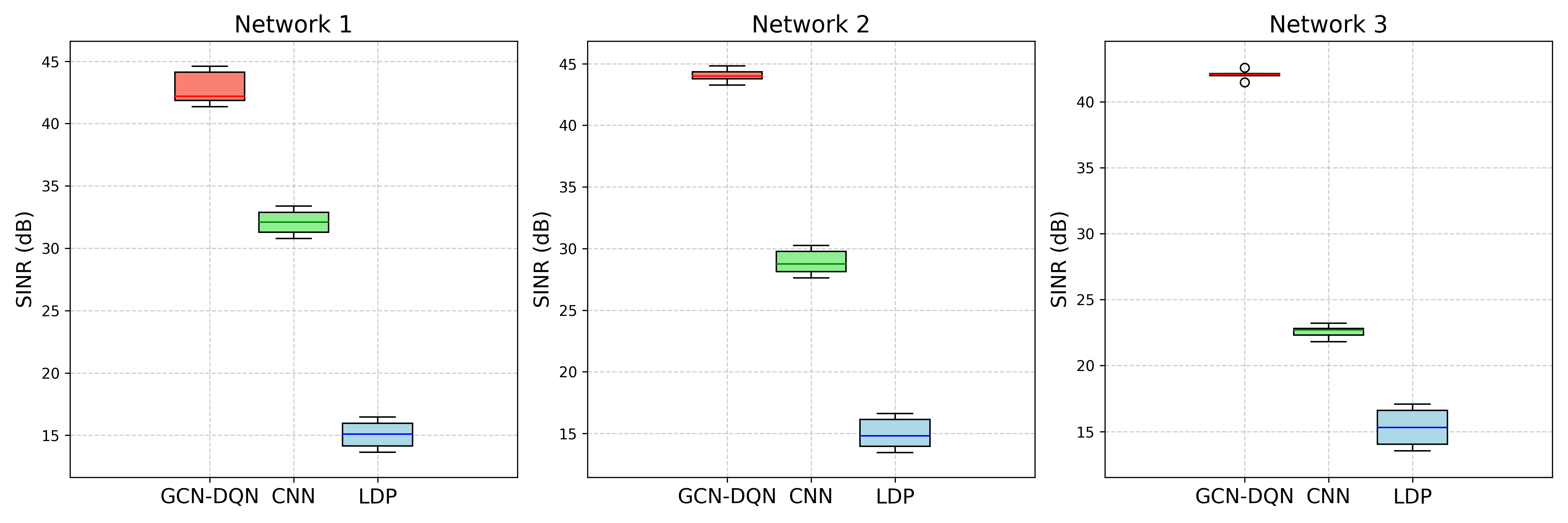}
\caption{Interference Effect on Receiver-Side SINR.}
\label{sinr_comparison}
\vspace{-10pt}
\end{figure}

Figure~\ref{sinr_comparison} shows the examined  receiver-side SINR performance of our GCN-DQN based method in comparison to both  the base LDP algorithm and our previous CNN-based LDP algorithm \cite{b14} .Our proposed GCN-DQN model demonstrates a significant improvement in SINR performance compared to both the CNN-based model and the LDP baseline. Specifically, it achieves a mean SINR of 42.20 dB, representing a 27.11 dB gain and 179.6\% improvement over the LDP approach. These results indicate a substantial enhancement in SINR with our GCN-DQN-based approach and highlight the model's ability to provide effective interference coordination, ensuring high network performance across varying network sizes. In terms of efficiency, the GCN-DQN model also offers faster inference compared to the CNN-based method, which has an inference time of 0.47 seconds. In contrast, the GCN-DQN model achieves significantly lower inference times of 0.0165 s, 0.0372 s, and 0.1107 s for networks with 83, 151, and 320 links, respectively. Moreover, as the network size approximately doubles, the inference time nearly triples, revealing a superlinear yet still practical growth trend that confirms the suitability of GCN-DQN for real-time scheduling scenarios.

\section{Conclusion}
This paper introduces a machine learning-based solution for interference coordination in multi-cell, multi-channel networks supporting large-scale, industrial URLLC applications with diverse real-time requirements. Initially, a CNN-based model was proposed in \cite{b14} to address inter-cell interference, leading to considerable improvements in SINR, network reliability, and real-time scheduling performance compared to the traditional LDP algorithm. Building upon this, we further proposed a Graph Convolutional Network with Deep Q-Learning (GCN-DQN) model that significantly outperforms both the CNN-based and LDP approaches across all network configurations.

Numerical results demonstrate that the GCN-DQN model outperforms both the baseline LDP algorithm and our previously proposed CNN-based method for interference coordination. Specifically, the GCN-DQN model achieves mean SINR improvements of 179.6\%, 197.4\%, and 175.2\% over the LDP approach across three network configurations. In comparison, the GCN-DQN model also shows mean SINR improvements of 31.5\%, 53.0\%, and 84.7\% over the CNN-based approach across the same configurations.

Furthermore, the GCN-DQN model achieves a significant reduction in inference time, with times of 0.0165 s, 0.0372 s, and 0.1107 s for the three configurations, respectively. As the network size increases, the inference time grows at a near tripling rate as the network size doubles, demonstrating scalability and real-time feasibility.

Our results underline the effectiveness of deep reinforcement learning in managing interference, achieving higher SINR, and improving network schedulability, especially in large-scale URLLC environments. The proposed GCN-DQN approach offers substantial improvements in both performance and computational efficiency, showcasing the transformative potential of graph-based machine learning in optimizing interference coordination and advancing the scalability and reliability of next-generation communication networks.

\vspace{12pt}
\color{red}

\end{document}